\documentclass[aps,twocolumn,showpacs,prl,floatfix,superscriptaddress]{revtex4}

\usepackage{epsfig}
\usepackage{graphics}
\usepackage{float}
\usepackage{lipsum}

\usepackage[T1]{fontenc}
\usepackage{mathptmx}

\begin{document}


\title{High-\textit{T}$_{c}$ superconductivity in clathrate calcium hydride CaH$_{6}$}

\author{Liang Ma}
\thanks{L.M., K.W. and Y.X. equally contributed to this work.}
\affiliation{State Key Laboratory of Superhard Materials, College of Physics, Jilin University, Changchun 130012, China}
\affiliation{International Center of Computational Method \& Software, College of Physics, Jilin University, Changchun 130012, China}
\affiliation{International Center of Future Science, Jilin University, Changchun 130012, China}

\author{Kui Wang}
\thanks{L.M., K.W. and Y.X. equally contributed to this work.}
\affiliation{State Key Laboratory of Superhard Materials, College of Physics, Jilin University, Changchun 130012, China}
\affiliation{International Center of Computational Method \& Software, College of Physics, Jilin University, Changchun 130012, China}

\author{Yu Xie}
\thanks{L.M., K.W. and Y.X. equally contributed to this work.}
\affiliation{State Key Laboratory of Superhard Materials, College of Physics, Jilin University, Changchun 130012, China}
\affiliation{International Center of Computational Method \& Software, College of Physics, Jilin University, Changchun 130012, China}
\affiliation{Key Laboratory of Physics and Technology for Advanced Batteries (Ministry of Education), Jilin University, Changchun 130012, China}

\author{Xin Yang}
\author{Yingying Wang}
\affiliation{State Key Laboratory of Superhard Materials, College of Physics, Jilin University, Changchun 130012, China}
\affiliation{International Center of Computational Method \& Software, College of Physics, Jilin University, Changchun 130012, China}

\author{Mi Zhou}
\affiliation{International Center of Computational Method \& Software, College of Physics, Jilin University, Changchun 130012, China}

\author{Hanyu Liu}
\affiliation{International Center of Computational Method \& Software, College of Physics, Jilin University, Changchun 130012, China}
\affiliation{International Center of Future Science, Jilin University, Changchun 130012, China}
\affiliation{State Key Laboratory of Superhard Materials and Key Laboratory of Physics and Technology for Advanced Batteries (Ministry of Education), College of Physics,  Jilin University, Changchun 130012, China}

\author{Xiaohui Yu}
\affiliation{Beijing National Laboratory for Condensed Matter Physics and Institute of Physics, Chinese Academy of Sciences, Beijing 100190, China}

\author{Yongsheng Zhao}
\affiliation{Deutsches Elektronen-Synchrotron DESY, Hamburg 22607, Germany}

\author{Hongbo Wang}
\email{whb2477@jlu.edu.cn}
\affiliation{State Key Laboratory of Superhard Materials, College of Physics, Jilin University, Changchun 130012, China}
\affiliation{International Center of Computational Method \& Software, College of Physics, Jilin University, Changchun 130012, China}

\author{Guangtao Liu}
\email{liuguangtao@jlu.edu.cn}
\affiliation{International Center of Computational Method \& Software, College of Physics, Jilin University, Changchun 130012, China}

\author{Yanming Ma}
\email{mym@jlu.edu.cn}
\affiliation{State Key Laboratory of Superhard Materials, College of Physics, Jilin University, Changchun 130012, China}
\affiliation{International Center of Computational Method \& Software, College of Physics, Jilin University, Changchun 130012, China}
\affiliation{International Center of Future Science, Jilin University, Changchun 130012, China}
\date{\today}

\begin{abstract}
Recent discovery of superconductive rare earth/actinide superhydrides has ushered in a new era of superconductivity research at high pressures. This distinct type of clathrate metal hydrides was first proposed for alkaline-earth-metal hydride CaH$_{6}$ that, however, has long eluded experimental synthesis, impeding an understanding of pertinent physics. Here, we report successful synthesis of CaH$_{6}$ and its measured superconducting critical temperature \textit{T}$_{c}$ of 215 K at 172 GPa, which is evidenced by a sharp drop of resistivity to zero and a characteristic decrease of \textit{T}$_{c}$ under a magnetic field up to 9 T. An estimate based on the Werthamer--Helfand--Hohenberg model gives a giant zero-temperature upper critical magnetic field of 203 T. These remarkable benchmark superconducting properties place CaH$_{6}$ among the most outstanding high-\textit{T$_{c}$} superhydrides, marking it as the hitherto only clathrate metal hydride outside the family of rare earth/actinide hydrides. This exceptional case raises great prospects of expanding the extraordinary class of high-\textit{T$_{c}$} superhydrides to a broader variety of compounds that possess more diverse material features and physics characteristics.
\end{abstract}


\maketitle

Superconductivity is a remarkable quantum phenomenon of broad interest for its great scientific significance and immense application potential \cite{A1,A2,A3,A4}. Both of these factors, especially the latter, hinges on the discovery of superconductors with critical transition temperature \textit{T$_{c}$} near room temperature that would allow wide-ranging device and equipment implementations ranging from high-precision magnetic signal detection, to super-strong magnets, and long-distance power delivery \cite{A2,A3}. While conventional Bardeen-Cooper-Schrieffer (BCS) theory predicts atomic metallic hydrogen could be the most plausible room-temperature superconductor because of its high Debye temperature and strong electron-phonon coupling \cite{A5,A6,A7,A8}, the realization of solid hydrogen is extremely challenging due to the required high pressures that push the current experimental limits \cite{A9,A10,A11}. Hydrogen-rich compounds were proposed as alternative candidate materials for exploring high-\textit{T$_{c}$} superconductivity since the chemical pre-compression induced by the incorporated elements that hold together hydrogen atoms in the compounds could significantly lower the metallization pressure of hydrogen to experimentally reachable regimes \cite{A12,A13}. Despite extensive earlier efforts \cite{A1,A2,A14,A15,A16}, the measured or predicted \textit{T$_{c}$} values of compressed hydrides stayed well below those of unconventional cuprate superconductors (up to 164 K at 31 GPa \cite{A17}). This situation was changed by the breakthrough discovery in 2012 when an advanced crystal structure search approach predicted a sodalite-like ionic clathrate calcium hydride CaH$_{6}$ with a remarkably high \textit{T$_{c}$} of 220-235 K at 150 GPa \cite{A18}. This work was followed by the discovery of superconductivity at 203 K in H$_{3}$S under the high pressure of 155 GPa in 2014 \cite{A19}. Notably, the high \textit{T$_{c}$} value of CaH$_{6}$ is rooted in the clathrate hydrogen framework that closely mimics the physical characteristics and properties of atomic metallic hydrogen, which offers a new design principle and general material platform to search and discover high-\textit{T$_{c}$} hydrogen-rich compounds known as superhydrides \cite{A2,A15}. Based on the same idea in constructing CaH$_{6}$, a distinct class of ionic clathrate MH$_{6}$ hydrides (M represents metal elements) and two new series of rare-earth hydrides REH$_{9}$ and REH$_{10}$ were later predicted to possess higher \textit{T$_{c}$} values approaching even above room temperature \cite{A20,A21}. Inspired by these findings, a long list of superconducting superhydrides, including YH$_{6}$ \cite{A22,A23}, YH$_{9}$ \cite{A23,A24}, CeH$_{9}$, CeH$_{10}$ \cite{A25}, ThH$_{9}$, ThH$_{10}$ \cite{A26}, LaH$_{10}$ \cite{A27,A28}, and (LaY)H$_{10}$ \cite{A29}, were successfully synthesized with observed \textit{T$_{c}$} values ranging from 57 to 262 K at high pressures that agree well with theoretical predictions, opening a new path toward finding room-temperature superconductors.

\begin{figure}[!t]
	\begin{center}
		\epsfxsize=8cm
		\epsffile{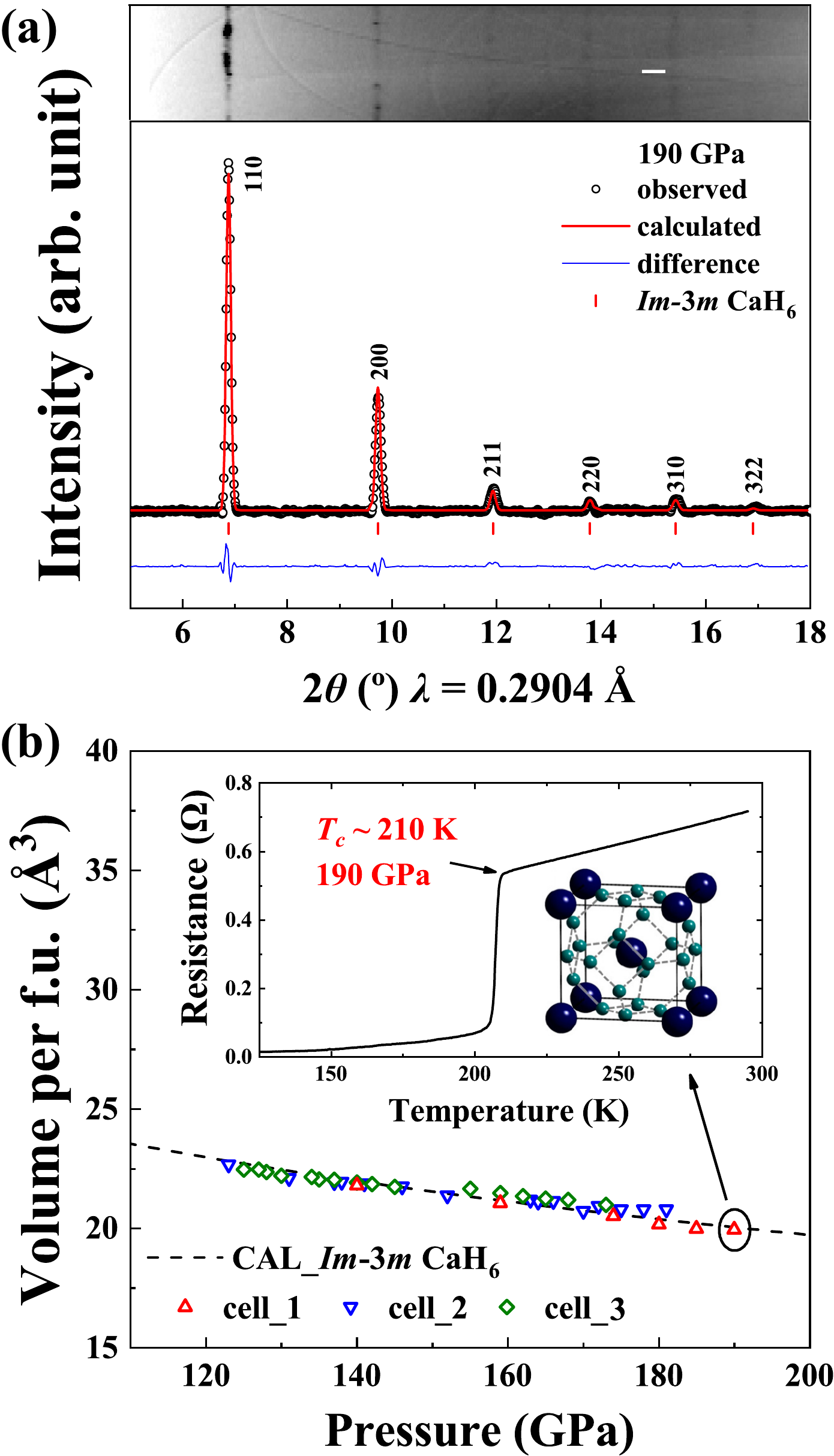}
	\end{center}
	\caption{ (a) Synchrotron X-ray diffraction pattern of the clathrate calcium hydride from cell\_1 obtained following laser heating of Ca and BH$_{3}$NH$_{3}$ at 190 GPa and the Rietveld refinement of the clathrate CaH$_{6}$ structure. (b) Experimental EOS from the different samples in this work in comparison with that of the predicted clathrate CaH$_{6}$. The EOS data from cell\_1, cell\_2, and cell\_3 were marked with red triangles, blue inverted triangles, and green rhombus, respectively. The superconducting transition with \textit{T$_{c}$} $\mathrm{\sim}$ 210 K was observed in the electrical measurement for cell\_1 at 190 GPa as shown in the top inset. }
	\label{fig:phase}
\end{figure}

Conspicuously, all the currently realized high-\textit{T$_{c}$} superhydrides contain rare earth (RE) or actinide (AC) elements as the anchor sites to provide electrons to bind and hold hydrogen atoms into the desired clathrate structure \cite{A1,A2,A15,A16}. This situation raises fundamental questions about whether this striking capacity can be found among the main-group and transition-metal elements for constructing a larger number and variety of high-\textit{T$_{c}$} superhydrides, which ignites the interest to reexamine the first predicted clathrate superhydride CaH$_{6}$ \cite{A18}, which has the lowest stabilization pressure of about 150 GPa among all the non-RE/AC metal clathrate superhydrides but has stayed out of reach despite repeated experimental attempts. In this work, we report successful synthesis of clathrate CaH$_{6}$ at pressures of 160-190 GPa and temperature of 2,000 K. X-ray diffraction (XRD) and equation of state (EOS) measurements confirm the theoretically predicted crystal structure. Ensuing electrical transport measurements revealed a superconducting transition temperature of 215 K at 172 GPa. These results not only finally verify and confirm the theoretical prediction made almost a decade ago, but more important, offer insights for exploring the hitherto still largely untapped main-group and transition-metal elements to serve as anchor sites in forming superhydrides that may exhibit more diverse structural and superconducting characteristics.

We have prepared 14 cells filled with a mixture of Ca foil and BH$_{3}$NH$_{3}$ as starting materials designated as cell\_1 to cell\_14. The samples were compressed at room temperature to 160-190 GPa and then heated to about 2,000 K with a one-sided pulsed radiation from a YAG laser. XRD measurements were conducted to determine the crystal structures of the products in cell\_1-3. Fig. 1a shows the XRD patterns of the product in cell\_1 at 190 GPa, which do not match any of the known calcium hydrides, such as CaH$_{2}$, CaH$_{4}$ \cite{A30,A31}, and Ca$_{2}$H$_{5}$ \cite{A30}. Instead, we find that these peaks can be indexed by a body-centered cubic (\textit{bcc}) lattice of space group \textit{Im}$\overline{\mathrm{3}}$\textit{m}, with a refined lattice parameter \textit{a} = 3.422 {\AA} (\textit{V} = 40.07 {\AA}$^{3}$). The volume of this \textit{bcc} structure agrees well with the previous theoretical prediction of clathrate CaH$_{6}$ (\textit{V} = 40.09 {\AA}$^{3}$ at 190 GPa) \cite{A18}. Although the occupation details of hydrogen atoms cannot be determined from the experiments due to the weak X-ray scattering cross section, the measured unit cell volumes can be used to estimate the stoichiometry of the hydrides. The \textit{bcc} unit cell consists of two Ca atoms occupying the atomic volume of 9.41 {\AA}$^{3}$/atom \cite{A32}, leaving 21.25 {\AA}$^{3}$ for hydrogen atoms. The volume occupied by each hydrogen atom was determined from the extrapolated equation of states (EOS) of H$_{2}$ to be roughly 1.78 {\AA}$^{3}$ \cite{A33}. On this basis, this experimentally synthesized new calcium hydride can be designated as CaH$_{5.97}$, which is in good agreement with the predicted stoichiometry of CaH$_{6}$. Furthermore, the EOS fitted in the unloading process (Fig. S1 \cite{A34}) is highly consistent with the theoretical EOS of CaH$_{6}$ as shown in Fig. 1b. All the evidence demonstrates that we have successfully synthesized the long-sought first predicted clathrate superhydride CaH$_{6}$. It is noteworthy that the diffraction pattern of clathrate CaH$_{6}$ was also observed in cell\_2 and cell\_3, as shown in Fig. S2 and S3 \cite{A34}, with the concurrent appearance of the previously reported CaH$_{4}$ \cite{A30,A31}.

\begin{figure}[!t]
	\begin{center}
		\epsfxsize=8cm
		\epsffile{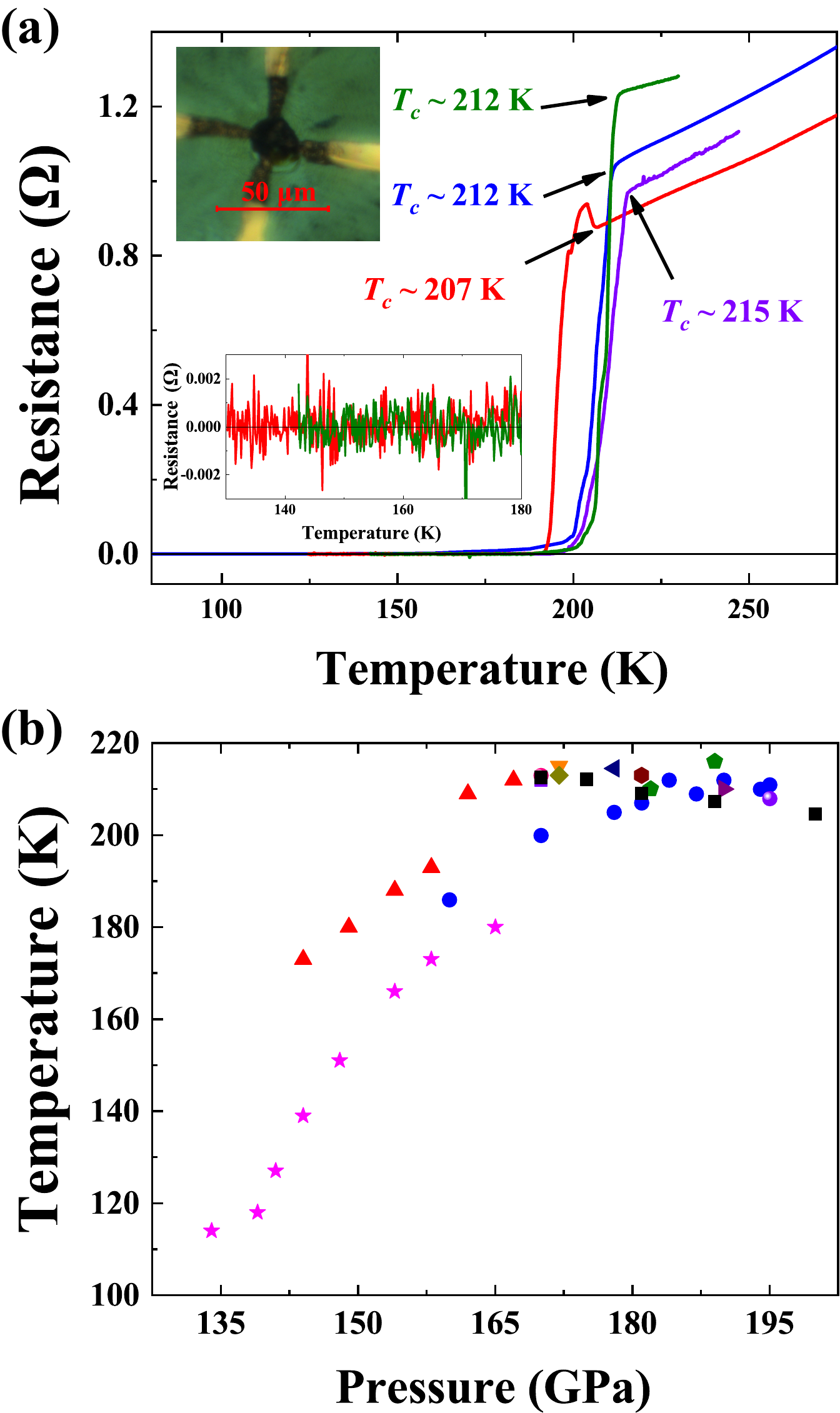}
	\end{center}
	\caption{ (a) Resistance measurements on the synthesized clathrate calcium hydride CaH$_{6}$. Optical micrograph of a sample at 170 GPa after laser heating is shown in the upper left panel inset. Red curve: sample (cell\_4) heated at 181 GPa with \textit{T$_{c}$} $\mathrm{\sim}$ 207 K; olive curve: sample (cell\_5) heated at 170 GPa with \textit{T$_{c}$} $\mathrm{\sim}$ 212 K; blue curve: sample (cell\_6) heated at 170 GPa with \textit{T$_{c}$} $\mathrm{\sim}$ 212 K; purple curve: sample (cell\_7) heated at 172 GPa with \textit{T$_{c}$} $\mathrm{\sim}$ 215 K. The resistance data with near zero values are shown on a smaller scale in the left bottom inset. (b) The dependence of the critical temperature \textit{T$_{c}$} on pressure (cell\_1, cell\_2, and cell\_4-14); the results from twelve different experiments are marked in different colors. The temperature dependence of the resistance at high pressures for the symbols of black square (cell\_6), red triangle (cell\_8) and purple star (cell\_9) are plotted in Fig. S4, S5a, and S5b \cite{A34}, respectively. }
	\label{fig:phase}
\end{figure}

Previous simulations have indicated that CaH$_{6}$ possess an estimated \textit{T}$_{c}$ of 220-235 K at 150 GPa \cite{A18}. To verify this result, we have\textbf{ }conducted electrical transport measurements. Representative electrical resistance measurements as a function of temperature at high pressures are shown in Fig. 2a, which clearly show superconducting transitions as evidenced by the sharp drop of the resistance at 215, 212, 207 K and 212 K at about 172, 170, 181 GPa and 170 GPa, respectively. Among of these experiments, zero resistance was observed on samples in cell\_4 and cell\_5 (inset in Fig. 2a), excluding the possibility that the abrupt drop of resistance on cooling arises from structural transitions. To determine the highest value of \textit{T$_{c}$}, we evaluated the pressure dependence of \textit{T$_{c}$} as shown in Fig. 2b. In different experimental runs, \textit{T$_{c}$} fluctuates slightly under the pressure of 170-190 GPa. The highest \textit{T$_{c}$} of 215 K at 172 GPa observed in the sample is consistent with our previous theoretical estimation of 213 K at this pressure (derived from the pressure coefficient of \textit{dT$_{c}$}/\textit{dP} $\mathrm{=-}$0.33 K/GPa) for clathrate CaH$_{6}$ \cite{A18}. Notably, we also conducted electrical measurements on samples in cell\_1 and cell\_2 before performing XRD measurements and observed a sharp drop of resistance around 210 K as shown in the inset of Fig. 1b and Fig. S3a \cite{A34}, respectively, indicating that the high-\textit{T$_{c}$} superconductivity indeed comes from clathrate CaH$_{6}$. Upon decompression, \textit{T$_{c}$} shows a dramatic drop below about 170 GPa as shown in Fig. 2b. It is interesting to observe that the pressure dependence of \textit{T$_{c}$} varies in different unloading experiments. This maybe caused by the different degrees of anisotropic stresses that are present during the decompression process, leading to variable distortions of the crystal lattice in different experiments. It is noted that distortion of crystal lattice has also been observed in clathrate LaH$_{10}$ \cite{A39}. Furthermore, for the sample in cell\_7, the decrease of pressure to about 130 GPa leads to the disappearance of superconducting transition as shown in Fig. S5a \cite{A34}, indicating possible decomposition of the superconducting phase. The anomalous resistance peak at about 200 K (the red curve in Fig. 1a), which was also observed in LaH$_{x}$ \cite{A28}, may be caused by the quantum confinement and coherence effects of inhomogeneous superconductivity in the presence of disorder \cite{A40}.

\begin{figure}[h]
	\begin{center}
		\epsfxsize=8.4cm
		\epsffile{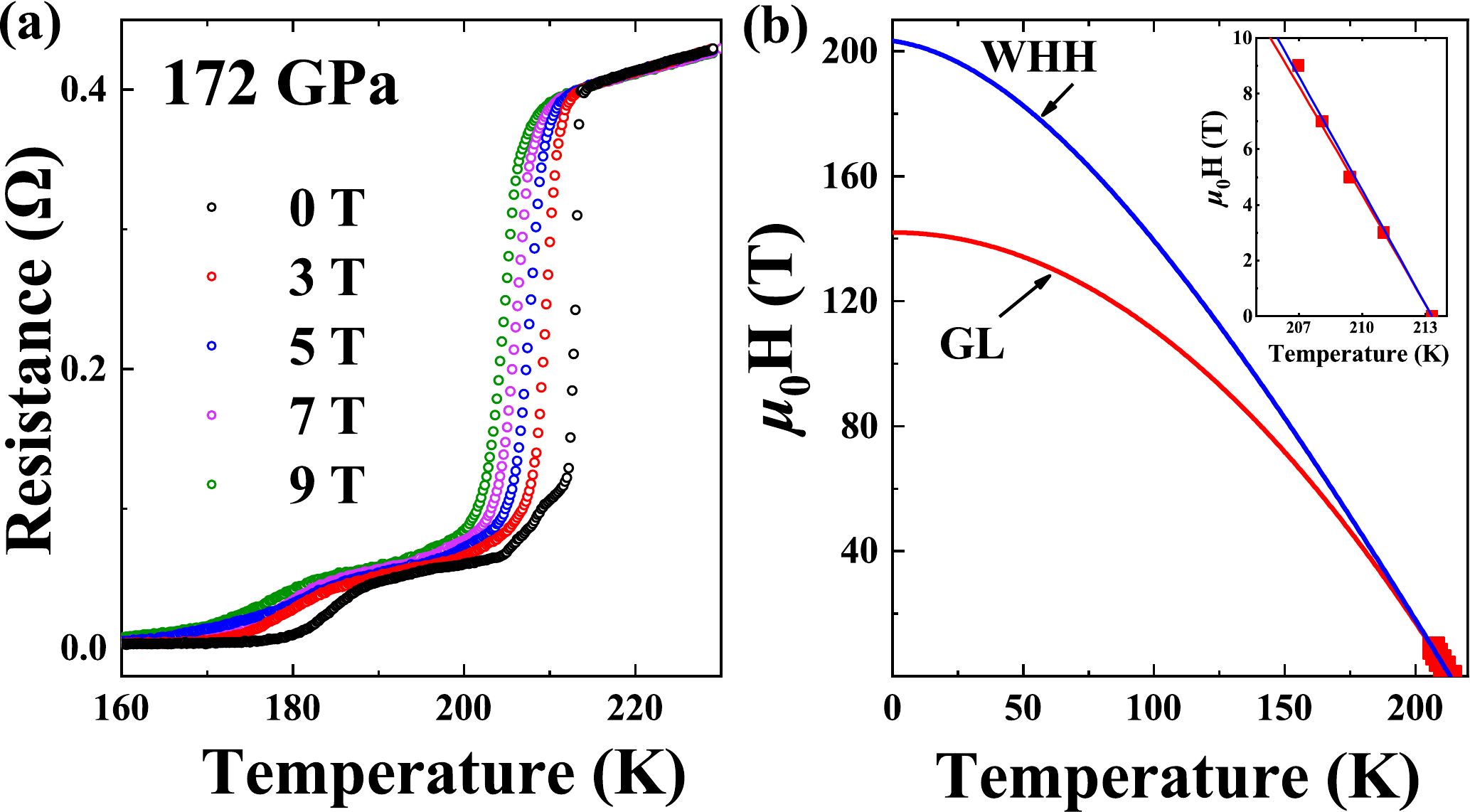}
	\end{center}
	\caption{(a) Temperature dependence of the electrical resistance under applied magnetic fields of H = 0, 3, 5, 7, and 9 T at 172 GPa. (b) Upper critical field \textit{H$_{c2}$} versus temperature following the criteria of 90\% of the resistance in the metallic state at 172 GPa, fitted with the GL and WHH models. Inset: the dependence of the \textit{T$_{c}$} under the applied magnetic field.}
	\label{fig:phase}
\end{figure}

The typical size of our calcium hydride samples is rather small (20-25 \textit{$\mu$}m in diameter), making it almost impossible to probe the weak signal of Meissner effect in the current experimental capabilities \cite{A24,A28}. Nevertheless, the superconducting nature of the transitions can be verified by its dependence on external magnetic fields. An applied external field could break the Cooper pairs due to the Pauli paramagnetic effect of electron spin polarization and the diamagnetic effect of the orbital motion, thus reducing the value of \textit{T$_{c}$}. As shown in Fig. 3a, the resistance drop gradually shifts to lower temperatures as the magnetic field is increased in the range 0-9 T at 172 GPa. The upper critical field as a function of temperature, which is defined as 90\% of the resistance, is shown in the inset of Fig. 3b. The application of a magnetic field reduces \textit{T}${}_{c}$ by about 6 K at \textit{$\mu$}$_{0}$\textit{H} = 9 T. The extrapolation values of the upper critical field \textit{$\mu$}$_{0}$\textit{H}$_{c2}$(T) and the coherence length towards T = 0 K are 142 T and 15.2 {\AA} and 203 T and 12.7 {\AA} fitted by the Ginzburg--Landau (GL) \cite{A41,A42} and Werthamer--Helfand--Hohenberg (WHH) \cite{A43} models, respectively. Another independent experiment (cell\_11) at 178 GPa (Fig. S6 \cite{A34}) shows similar results, producing the \textit{$\mu$}$_{0}$\textit{H}$_{c2}$(0) and the coherence length estimated with GL and WHH models at 132 T and 15.8 {\AA} and 181 T and 13.5 {\AA}, respectively. The magnitude of the estimated coherence lengths ($\mathrm{\sim}$13-16 {\AA}) is well below the typical value (hundreds of angstroms) for type-I superconductors, indicating the CaH$_{6}$ is a strongly type-II superconductor.

Past studies have attempted to synthesize the predicted high-\textit{T$_{c}$} clathrate calcium superhydride CaH$_{6}$ but without success \cite{A30,A31}. One of the probable reasons for the failure of previous experiments is that Ca and pure H$_{2}$ were used as precursors, whereas Ca easily reacts with H$_{2}$ at low pressures to form low-hydrogen-content CaH$_{x}$, such as CaH$_{2}$, which is hard to further react with H$_{2}$ [30]. A viable solution is to find hydrogen sources that only release H$_{2}$ at preferred conditions for CaH$_{6}$ synthesis. Adopting this idea, BH$_{3}$NH$_{3}$, which releases H$_{2}$ at higher temperatures, was selected as the hydrogen source in our present work, leading to the successful synthesis of clathrate CaH$_{6}$. After the completion of this work, we became aware of another independent experiment by Li \textit{et al}. \cite{A44} where similar \textit{T}$_{c}$ was also observed in synthesized calcium superhydride using BH$_{3}$NH$_{3}$ as the H$_{2}$ source.

In summary, we have successfully synthesized the first predicted and long-sought sodalite-like clathrate calcium superhydride CaH$_{6}$ that exhibits a superconducting transition temperature of 215 K at 172 GPa, which represents the highest \textit{T$_{c}$} value among non-RE ionic superhydrides. This result confirms the original theoretical prediction and provides impetus for further exploration of high-\textit{T$_{c}$} clathrate compounds. The present findings are expected to greatly expand the scope of ongoing studies in search of room-temperature superconductors among more diverse material classes.

${}$

We are grateful to Prof. Zhongxian Zhao for his kind suggestions and help in many aspects of this work. This work was supported by the Major Program of the National Natural Science Foundation of China (Grant No. 52090024), the Strategic Priority Research Program of Chinese Academy of Sciences (Grant No. XDB33000000), National Key R\&D Program of China (Grant no. 2018YFA0305900), National Natural Science Foundation of China (Grant No. 11874175, 12074139, 12074138, 11874176, 12034009 and 11974134), Jilin Province Outstanding Young Talents Project (Grant No. 20190103040JH), and Program for JLU Science and Technology Innovative Research Team (JLUSTIRT). XRD measurements were performed at P02.2 station of PETRA III in DESY (Hamburg, Germany), BL15U1 station in SSRF (Shanghai, China), 4W2 station in BSRF (Beijing, China) and BL10XU in SPring-8 (Sayo, Japan). The measurements of superconducting transition under external magnetic field were supported by the SECUF (Beijing, China) and SHMFF (Hefei, China).

\end{document}


\title{Supplementary Material for\\ ``High-\textit{T}$_{c}$ superconductivity in clathrate calcium hydride CaH$_{6}$''}

\author{{ Liang Ma,$^{1,2,3,\star}$ Kui Wang,$^{1,2,\star}$ Yu Xie,$^{1,2,4,\star}$ Xin Yang,$^{1,2}$ Yingying Wang,$^{1,2}$ Mi Zhou,$^{2}$ Hanyu Liu,$^{2,3,5}$ Xiaohui Yu,$^{6}$ Yongsheng Zhao,$^{7}$ Hongbo Wang,$^{1,2,\dagger}$, Guangtao Liu,$^{2,\ddagger}$ and Yanming Ma$^{1,2,3,\S}$}\\
{\small \em $^1$State Key Laboratory of Superhard Materials, College of Physics, Jilin University, Changchun 130012, China\\
$^2$International Center of Computational Method \& Software, College of Physics, Jilin University, Changchun 130012, China\\
$^3$International Center of Future Science, Jilin University, Changchun 130012, China\\
$^4$Key Laboratory of Physics and Technology for Advanced Batteries (Ministry of Education), Jilin University, Changchun 130012, China\\
$^5$State Key Laboratory of Superhard Materials and Key Laboratory of Physics and Technology for Advanced Batteries (Ministry of Education), College of Physics, Jilin University, Changchun 130012, China\\
$^6$Beijing National Laboratory for Condensed Matter Physics and Institute of Physics, Chinese Academy of Sciences, Beijing 100190, China\\
$^7$Deutsches Elektronen-Synchrotron DESY,Hamburg 22607, Germany\\}
{\small $^{\dagger}$Electronic Address: whb2477@jlu.edu.cn\\
$^{\ddagger}$Electronic Address: liuguangtao@jlu.edu.cn\\
$^{\S}$Electronic Address: mym@jlu.edu.cn\\
$^{\star}$ L.M., K.W. and Y .X. equally contributed to this work}}

\date{\today}

\maketitle

\newpage

\begin{center}
	\textbf{II. METHODS}
\end{center}

We synthesized calcium hydride via a reaction of calcium (Alfa Aesar 99.5\%) and BH$_{3}$NH$_{3}$ (Sigma-Aldrich, 97\%) in a diamond anvil cell (DAC). The diamonds used in DACs had culets with diameters of 50-60 \textit{$\mu$}m and were beveled at 8.5$^{o}$ to diameter of about 250 \textit{$\mu$}m. Composite gasket consisting of a rhenium outer annulus and a cubic boron nitride (\textit{c}-BN) epoxy mixture insert was employed to contain the sample while isolating the electrical leads in the electrical measurements. Ca foil with the thickness of 2 \textit{$\mu$}m was sandwiched between the BH3NH3. Sample preparation was done in an inert Ar atmosphere with residual O$_{2}$ and H$_{2}$O contents of < 0.01 ppm to guarantee that the sample was properly isolated from the surrounding atmosphere. Afterwards, the samples were compressed to required pressures at room temperature. The pressure values in cell\_1 and cell\_2 was determined by the EOS of contacted Au electrode \cite{A1}, while pressures in the other cells were determined from room-temperature diamond Raman measurements \cite{A2}. One-side laser-heating experiments were performed using a pulsed YAG laser. The temperature was determined using the emission spectra of the blackbody radiation within the Planck’s radiation law. \textit{In situ} high-pressure powder XRD experiments with cell\_1 at room temperature were performed at beamline DESY (Hamburg), using the flat panel detector XRD1621 from PerkinElmer. The X-ray wavelength was ${\lambda}$ = 0.2904 {\AA}, and the beam size was 1.9 ${\times}$ 2.2 \textit{$\mu$}m$^{2}$ at the full width at half maximum (FWHM). The XRD patterns of cell\_2 and cell\_3 were obtained principally at beamline BL15U1 (${\lambda}$ = 0.6199 {\AA}) of Shanghai Synchrotron Radiation Facility using a focused (5 ${\times}$ 12 \textit{$\mu$}m$^{2}$) monochromatic X-ray beam. A Mar165 CCD was used as the detector. Additional experiments were conducted at the 4W2 High Pressure Station of the Beijing Synchrotron Radiation Facility and BL10XU in SPring-8. The sample to detector distance and other geometric parameters were calibrated using a CeO$_{2}$ standard. The software package Dioptas was used to integrate powder diffraction rings and convert the 2-dimensional data to 1-dimensional profiles \cite{A3}. The full profile analysis of the diffraction patterns and the Rietveld refinements were done using GSAS and EXPGUI packages \cite{A4}.

The resistances were all measured via the four-probe van der Pauw method where four Au electrodes were placed on the BH3NH3 plate with currents of 1-100 \textit{$\mu$}A. The temperature dependence of the electrical resistance was measured upon cooling and warming cycles with a slow temperature ratio (0.2 K·min$^{-1}$). The data were taken upon warming as it yields a more accurate temperature reading. Cu-Be alloy DACs were used for magnetotransport measurements under external magnetic fields up to 9 T. 

\newpage

\begin{center}
	\epsfxsize=14cm
	\epsffile{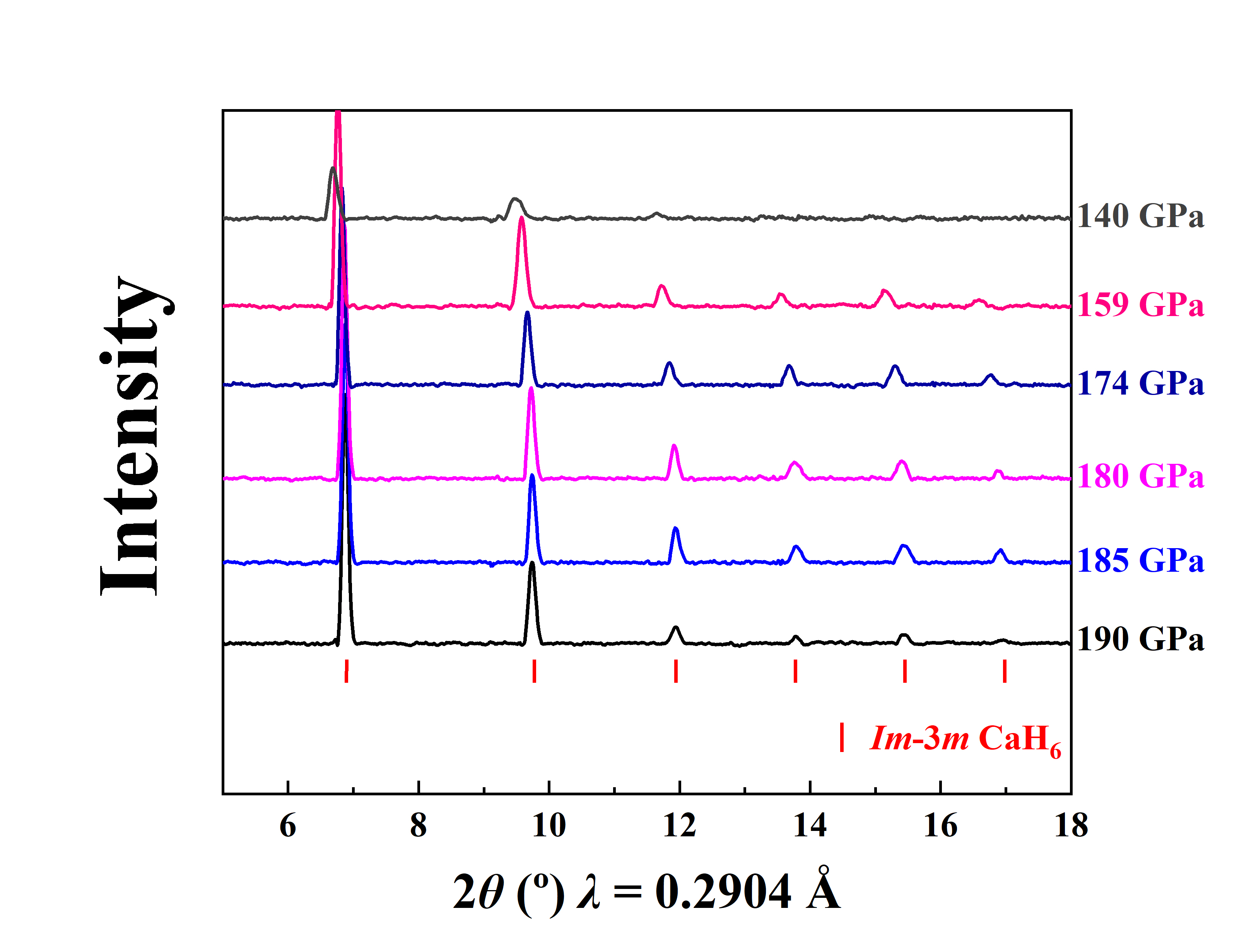}
\end{center}
\textbf{Fig. S1.} Synchrotron X-ray diffraction pattern of superconducting calcium hydride obtained following laser heating of Ca and BH$_{3}$NH$_{3}$ at 190 GPa (cell\_1) during decompression in the pressure range of 190–140 GPa. The red vertical lines mark the diffraction peak positions from \textit{Im}$\overline{\mathrm{3}}$\textit{m} CaH$_{6}$.
\label{fig:phase}

\begin{center}
	\epsfxsize=16cm
	\epsffile{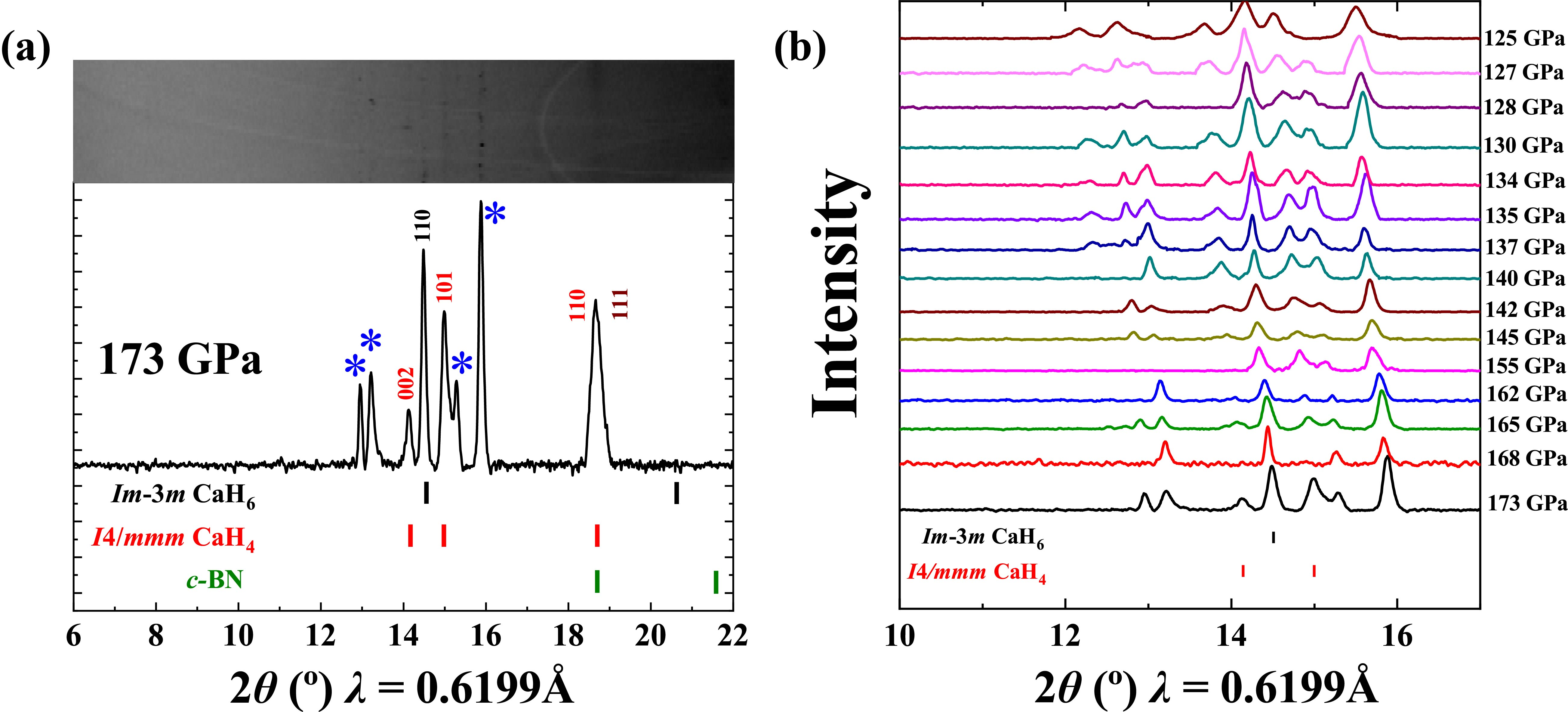}
\end{center}	
\textbf{FIG. S2.} (a) Synchrotron X-ray diffraction pattern of superconducting calcium hydrides obtained following laser heating of Ca and BH$_{3}$NH$_{3}$ at 173 GPa (cell\_3). The diffraction peak marked by the black vertical lines is consistent with the (110) reflection of \textit{Im}$\overline{\mathrm{3}}$\textit{m} CaH$_{6}$. The diffraction peaks marked by the red vertical lines are consistent with the (002) and (101) reflections of \textit{I}4/\textit{mmm} CaH$_{4}$. The diffraction peaks marked by the green vertical lines are from \textit{c}-BN gasket. Unidentified reflections are indicated by asterisks. (b) Experimental XRD patterns (cell\_3) during decompression in the pressure range of 173–125 GPa.
\label{fig:phase}

\begin{center}
	\epsfxsize=16cm
	\epsffile{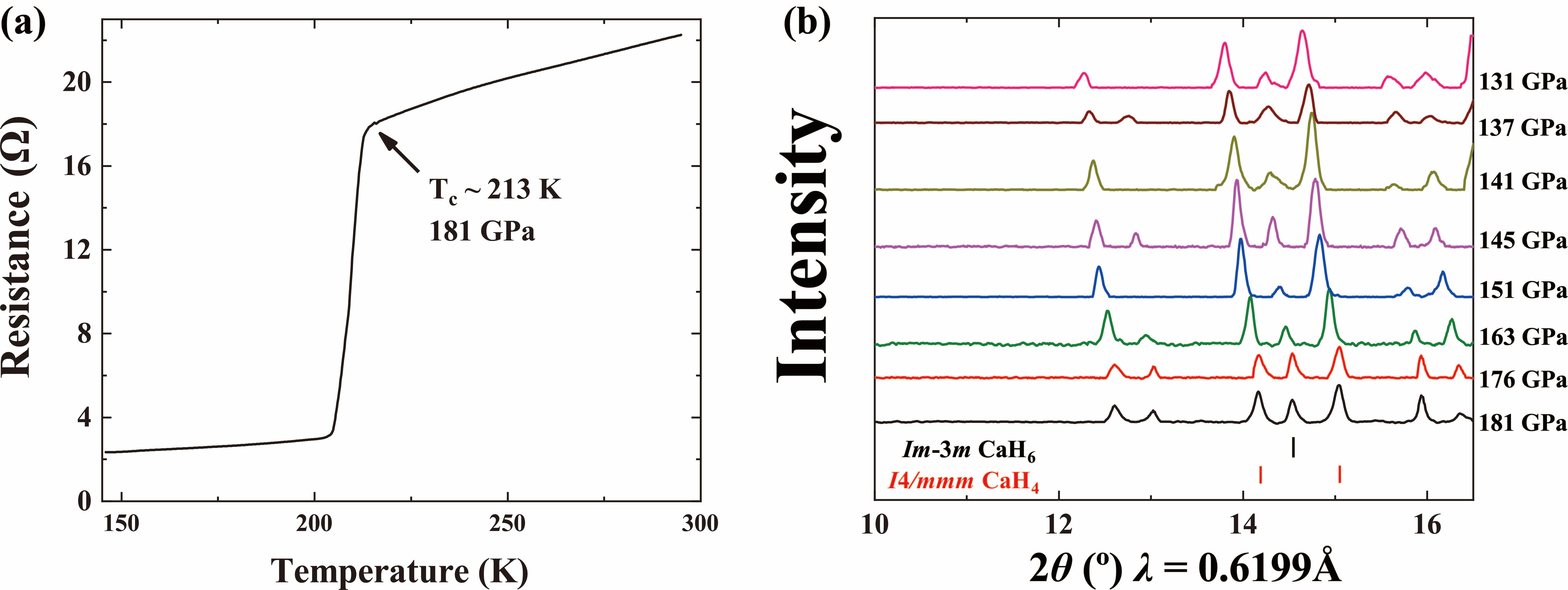}
\end{center}	
\textbf{FIG. S3.} (a) The superconducting transition at ~213 K was observed at 181 GPa (cell\_2). (b) Experimental XRD patterns (cell\_2) during decompression in the pressure range of 181–131 GPa. 
\label{fig:phase}

\begin{center}
	\epsfxsize=16cm
	\epsffile{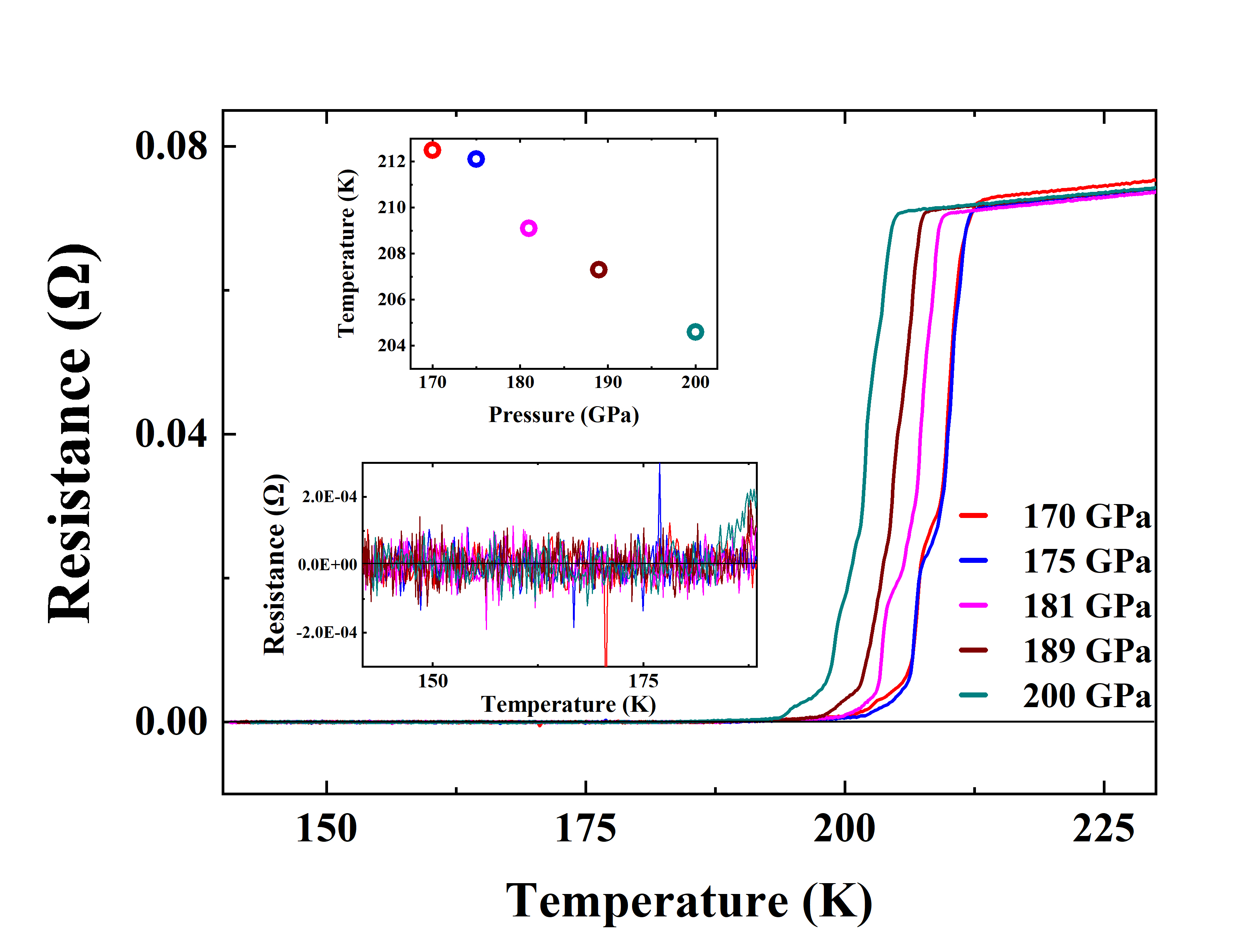}
\end{center}	
\textbf{FIG. S4.} Temperature dependence of resistance of the sample from cell\_4 measured upon compression. The sample was synthesized at around 170 GPa from a mixture of Ca and BH$_{3}$NH$_{3}$. The pressure dependence of the onset of superconductivity is shown in the inset.
\label{fig:phase}

\begin{center}
	\epsfxsize=16cm
	\epsffile{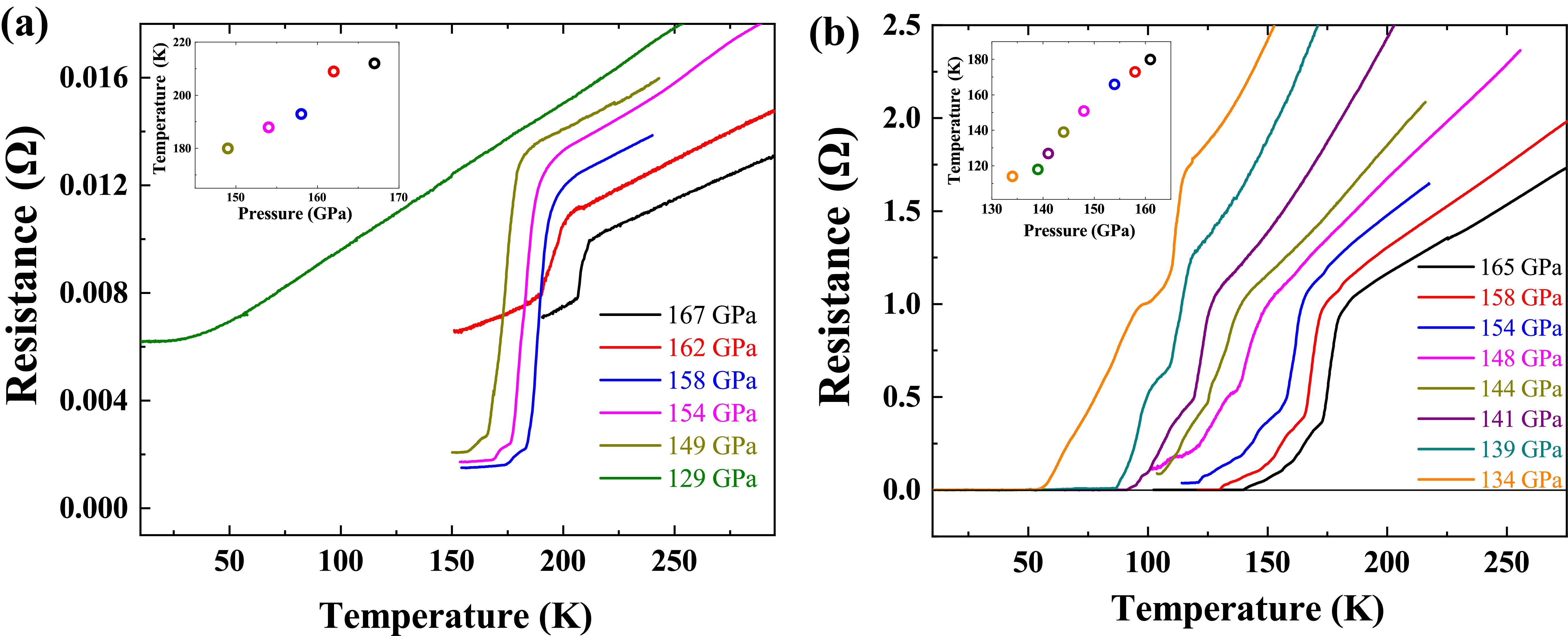}
\end{center}	
\textbf{FIG. S5.} Temperature dependence of resistance of the sample from cell\_7 and cell\_8 measured upon decompressions. (a) The sample (cell\_7) was synthesized at around 170 GPa from a mixture of Ca and BH$_{3}$NH$_{3}$. After heating, the pressure dropped to 167 GPa. (b) The sample (cell\_8) was synthesized at around 160 GPa from a mixture of Ca and BH$_{3}$NH$_{3}$. After heating, the pressure increased to 165 GPa. The pressure dependence of the onset of superconductivity is shown in the inset. 
\label{fig:phase}

\begin{center}
	\epsfxsize=16cm
	\epsffile{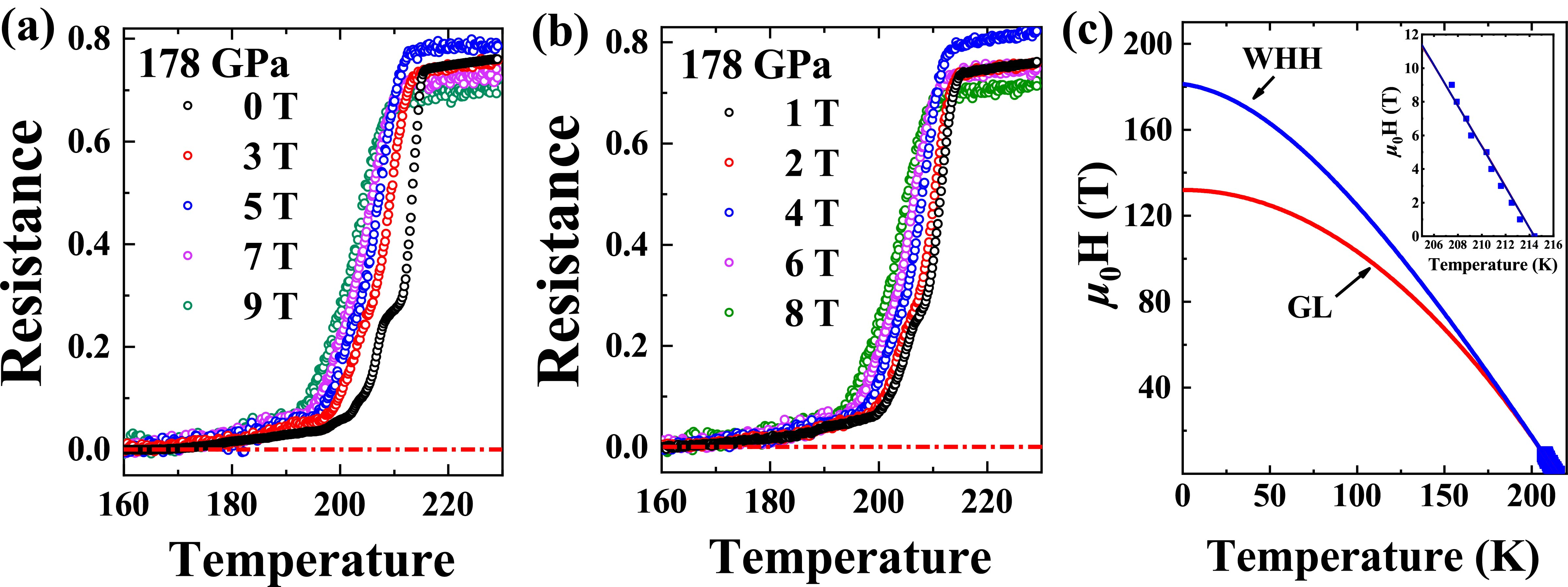}
\end{center}	
\textbf{FIG. S6.} Superconducting transitions of the calcium superhydride (cell\_11) at 178 GPa under different magnetic fields. (a) Temperature dependence of the resistance on the external magnetic fields (0, 3, 5, 7, 9 T); (b) Temperature dependence of the resistance on the external magnetic fields (1, 2, 4, 6, 8 T). To display clearly the suppression of the superconducting transition by the magnetic field, we show our data in two figures. (c) Upper critical field \textit{$\mu$}$_{0}$\textit{H}$_{c2}$(T) versus temperature following the criteria of 90\% of the resistance in the metallic state at 178 GPa, fitted with the GL and WHH models. Inset: the dependence of the High-\textit{T}$_{c}$ under the applied magnetic field. 
\label{fig:phase}

\begin{center}
	\epsfxsize=14cm
	\epsffile{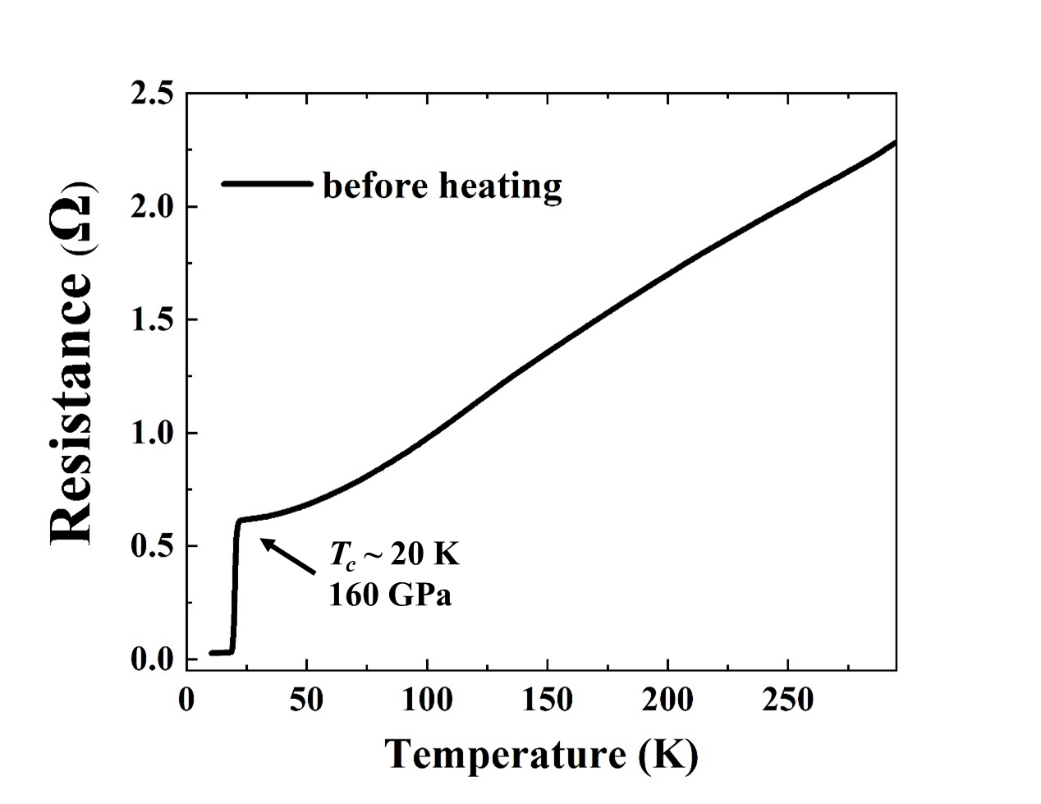}
\end{center}	
\textbf{FIG. S7.} Temperature dependence of the resistance in the warming cycle at around 160 GPa before laser heating. Superconductivity with a \textit{T}$_{c}$ ~20 K is consistent with that of elemental Ca solid. 
\label{fig:phase}